 \documentclass[final,5p,times,twocolumn]{elsarticle}

\usepackage{amssymb}
\usepackage{bm}
\usepackage{xcolor,soul}%
\usepackage{graphicx}
\biboptions{numbers,sort&compress}
\usepackage{framed}
\usepackage{nomencl}
\makenomenclature

\usepackage{multicol}
\renewcommand*\nompreamble{\begin{multicols}{2}}

\renewcommand*\nompostamble{\end{multicols}}

\usepackage[breaklinks=true]{hyperref}

\hypersetup{
    colorlinks,
    citecolor=black,
    filecolor=black,
    linkcolor=black,
    urlcolor=black
}
\usepackage{flushend}

\journal{International Journal of Heat and Mass Transfer}

\begin{document}

\begin{frontmatter}



\title{Thermal flux in unsteady Rayleigh-B\'{e}nard magnetoconvection}


%
\author[label1]{Sandip Das}
\author[label1]{ Krishna Kumar\corref{cor1}}%
 \ead{kumar.phy.iitkgp@gmail.com}
 \cortext[cor1]{Corresponding author}
\address[label1]{ 
Department of Physics, Indian Institute of Technology Kharagpur, Kharagpur-721302, India\\
}
\begin{abstract}
We present results of numerical investigation on thermal flux in Rayleigh-B\'{e}nard magnetoconvection in the presence of a uniform vertical magnetic field. We have studied thermal flux in different viscous fluids with a range of  Prandtl number ($0.1 \le \mathrm{Pr} < 6.5$) and a range of Chandrasekhar number ($50 \le \mathrm{Q} \le 2.5 \times 10^4$). The power spectral density of the Nusselt number varies with frequency $f$  approximately as $f^{-2}$. The probability distribution function of the fluctuating part of the Nusselt number is nearly normal distribution with slight asymmetric tails. For a fixed value the Rayleigh number $\mathrm{Ra}$, the time averaged Nusselt number $\langle \mathrm{Nu} (\mathrm{Q})\rangle$ decreases logarithmically with Chandrasekhar number for $\mathrm{Q} > \mathrm{Q}_c$, which depends on $\mathrm{Ra}$ and $\mathrm{Pr}$. The reduced Nusselt number  $\mathrm{Nu_r}$ $=$ $\langle \mathrm{Nu}(\mathrm{Q})\rangle/{\langle \mathrm{Nu}(0)\rangle}$ rises sharply, reaches a maximum slightly above unity and then start decreasing very slowly to unity as the value of a dimensionless parameter $\sqrt{\mathrm{Ra/(Q~Pr)}}$ is raised. The probability distribution function of the local thermal flux in the vertical direction is found to be asymmetric and non-Gaussian with a cusp at its maximum.
\end{abstract}

\begin{keyword}
Magnetoconvection, Rayleigh number, Chandrasekhar number, Nusselt number, local heat flux,  thermal boundary layer, Power spectral density, nanofluids 
\end{keyword}

\end{frontmatter}

\section{\label{sec:Intro}Introduction}

The  understanding of heat flux in magnetoconvective flows is a topic of intense research due to its potential industrial applications in nanofluids~\cite{kakac_pramuanjaroenkij_2009,khanafer_aithal_2013,
selimefendigil_oztop_2014}, biofluids~\cite{shahcheraghi_etal_2002}, electro-chemical process~\cite{waskaas_kharkats_1999} and material processing research~\cite{series_hurle_1991,waskaas_kharkats_1999,davidson_1999} in addition to its relevance in traditional areas like geophysics~\cite{olson_glatzmaier_1996,glazmaier_etal_nature_1999,marshall_schott_1999,busse_pesch_2006,
roberts_king_rep_prog_phys_2013} and astrophysics~\cite{bec_etal_1996,cattaneo_etal_2003,thompson_dalgaard_2003,ryu_etal_2008}. 
A thermally stratified system,  where a thin horizontal layer of a fluid is subjected to an adverse temperature gradient and simultaneously subjected to a uniform magnetic field, is known as Rayleigh-B\'{e}nard magnetoconvection (RBM)~\cite{chandrasekhar_1961,fauve_etal_1984,weiss_proctor_2014,basak_etal_2014}.
Chandrasekhar~\cite{chandrasekhar_1961} analysed the linear problem of thermal convection in a homogeneous fluid. He showed that a uniform vertical magnetic field delays the convective flow. In addition, he showed that the onset of convection is always stationary if $\mathrm{Pr}$ is greater than $\mathrm{Pm}$.

The dynamics of RBM is governed by four dimensionless quantities: \\
(1) Rayleigh number $\mathrm{Ra}$, which is the relative measure of the buoyancy force over the dissipative force,\\
(2) Chandrasekhar's number $\mathrm{Q}$, which is a measure of the strength of the Lorentz force, \\
(3) the thermal Prandtl number $\mathrm{Pr} = \nu/\kappa$  is a ratio of the effective kinematic viscosity $\nu$ and the effective thermal diffusivity $\kappa$,  and \\
(4) the effective magnetic Prandtl number $\mathrm{Pm}=\sigma \mu_0 \nu$, where $\sigma$ is the electrical conductivity of the fluid and $\mu_0$ is  the magnetic permeability of air. The magnetic diffusivity of the fluid is defined as $\eta = 1/(\mu_0 \sigma)$. 

Experiments on the measurement of thermal flux in magnetoconvection in metallic fluids~\cite{cioni_etal_2000,aurnou_olson_2001,burr_mueller_2001} showed that the transport of heat across the fluid layer in turbulent magnetoconvection was reduced significantly and the fluid flow was affected~\cite{basak_etal_2014}. It was also found that  the  time averaged Nusselt number $\langle \mathrm{Nu}\rangle$ showed scaling behaviour with Rayleigh number $\mathrm{Ra}$~\cite{cioni_etal_2000,aurnou_olson_2001}. The scaling exponent was found to depend on the strength of the applied magnetic field. A uniform nanofluid consists of homogeneous suspension of metallic  nanoparticles in an ordinary fluid, which is also known as a  base/carrier fluid.  The viscous, thermal, electrical and magnetic properties of a nanofluid depend on the properties of the base fluid as well as the properties of suspended nanoparticles.  There is hardly any work on the role of magnetic field on the heat flux in unsteady flows in fluids including nanofluids, liquid crystals and metallic fluids. 

\begin{table*}[!t]   
\begin{framed}
 \nomenclature{${\rho}_{p}$}{density of nanoparticles}
 \nomenclature{${\rho}_{f}$}{density of base fluid}
 \nomenclature{${\rho}$}{effective density of nanofluid}
 \nomenclature{$\nu$}{effective kinematic viscosity of nanofluid}
 \nomenclature{$\kappa$}{effective thermal diffusivity of nanofluid}
 \nomenclature{${\sigma}_{p}$}{electrical conductivity of nanoparticles}
 \nomenclature{${\sigma}_{f}$}{electrical conductivity of base fluid}
 \nomenclature{${\sigma}$}{effective electrical conductivity of nanofluid}
 \nomenclature{$\eta$}{effective magnetic diffusivity of nanofluid}
 \nomenclature{$\mu_0$}{magnetic permeability of air}
 \nomenclature{$\mu_{f}$}{dynamic viscosity of base fluid}
 \nomenclature{$\mu$}{effective dynamic viscosity of nanofluid}
 \nomenclature{$\alpha$}{thermal expansion coefficient of nanofluid}
 \nomenclature{$c_V$}{specific heat of nanofluid at constant volume}
 \nomenclature{$K_p$}{thermal conductivity  of nanoparticles}
 \nomenclature{$K_f$}{thermal conductivity  of base fluid}
 \nomenclature{$K$}{effective thermal conductivity of nanofluid}
 \nomenclature{$\beta$}{temperature gradient across the fluid layer}
 \nomenclature{$g$}{acceleration due to gravity}
 \nomenclature{$\phi$}{volume fraction of suspended nanoparticles}
 \nomenclature{$\mathrm{Pr}$}{effective thermal Prandtl number of nanofluid}
 \nomenclature{$\mathrm{Pm}$}{effective magnetic Prandtl number of nanofluid}
 \nomenclature{$\mathrm{Ra}$}{Rayleigh number}
 \nomenclature{$\mathrm{Q}$}{Chandrasekhar's number}
 \nomenclature{$\mathrm{Nu}$}{Nusselt number} 
 \nomenclature{$\sigma({\mathrm{Nu}})$}{standard deviation in Nusselt number} 
 \nomenclature{$\delta_{th}$}{thickness of the thermal boundary layer}
 \nomenclature{$\gamma$}{scaling exponent for $\delta_{th}$} 
\printnomenclature
\end{framed}
\end{table*}

In this article we present results of numerical simulations on both global  and local heat fluxes in RBM  with a uniform vertical magnetic field in water based nanofluids with low dilution of non-magnetic spherical nanoparticles.  We have computed Nusselt number $\mathrm{Nu(Q)}$, which is a measure of the global heat flux for non-zero value of Chandrasekhar's number $\mathrm{Q}$, which is a ratio of the Lorentz force per unit volume to the drag force due to magneto-viscous effect. The time averaged Nusselt number $\langle \mathrm{Nu(Q)\rangle}$ increases  slowly with Chandrasekhar number $\mathrm{Q}$ for smaller values of $\mathrm{Q}$.  The effective Prandtl number of fluid is varied from $0.1 \le \mathrm{Pr} \le 6.4$. As soon as $\mathrm{Q}$ is raised above a critical value $\mathrm{Q}_c$, which depends on $\mathrm{Ra}$ and $\mathrm{Pr}$,  $\langle \mathrm{Nu(Q)}\rangle$ decreases logarithmically with $\mathrm{Q}$ for a fixed value of $\mathrm{Ra}$. We have also plotted the variation of the reduced Nusselt number $\mathrm{Nu_r}$ $=$ $\langle \mathrm{Nu}(\mathrm{Q})\rangle/{\langle \mathrm{Nu}(0)\rangle}$ with a dimensionless parameter $\sqrt{\mathrm{Ra/(Q Pr)}}$, which is a ratio of the buoyancy and Lorentz forces. For fluids with Prandtl number $\mathrm{Pr} \le 4.0$, $\mathrm{Nu_r}$ increases sharply with the dimensionless parameter $\sqrt{\mathrm{Ra/(Q Pr)}}$. It attains a maximum slightly above unity and then begins decreasing slowly towards unity, as $\sqrt{\mathrm{Ra/(Q Pr)}}$ is further raised. The probability distribution of fluctuations in the Nusselt number is close to normal with slightly asymmetric tails. The probability distribution functions (PDF) of the local heat fluxes in the vertical direction are found to be non-Gaussian with a cusp at their maxima. PDFs are asymmetric about their maxima and have exponential tails.  

\section{\label{sec:system}Hydromagnetic System}
\noindent
We consider a thin horizontal layer of a homogeneous nano-fluid of effective density $\rho$ and thickness $d$, effective thermal expansion coefficient $\alpha$ and effective electrical conductivity $\sigma$ and  subjected to an adverse temperature gradient $\beta$ in the presence of a uniform magnetic field ${\bm{\mathrm{B}}}_0 = B_0 \bm{\mathrm{e}}_3$ directed along the vertical direction. Here $\bm{\mathrm{e}}_3$ is a unit vector in the vertically upward direction. The effective density $\rho$ and the electrical conductivity $\sigma$  are expressed~\cite{selimefendigil_oztop_2014} as:  
\begin{eqnarray}
\rho &=& (1-\phi) \rho_f + \phi \rho_p,\\
\sigma &=& (1-\phi) \sigma_f +  \phi \sigma_p,
\end{eqnarray} 
where $\phi$ is the volume fraction of the suspended spherically shaped nanoparticles of density  ${\rho}_{p}$ and electrical conductivity ${\sigma}_{p}$ in a base fluid of density ${\rho}_f$ and electrical conductivity $\sigma_f$. We may express the products $\rho \alpha$ and $\rho c_V$ for nanofluids~\cite{selimefendigil_oztop_2014} as:
\begin{eqnarray}
(\rho \alpha) &=& (1-\phi)(\rho \alpha)_{f} + \phi (\rho \alpha)_{p},\\
(\rho c_V) &=& (1-\phi)(\rho c_V)_{f} + \phi (\rho c_V)_{p},
\end{eqnarray} 
where $c_V$ stands for the effective specific heat of nanofluid at constant volume. The effective thermal conductivity $K$ of a nanofluid~\cite{maxwell_1873} with spherical nanoparticles of thermal conductivity $K_p$ in a base fluid of thermal conductivity $K_f$ is expressed as
\begin{equation}
K = K_{f} \left[ \frac{(K_{p}+2K_{f}) - 2\phi(K_{f}-K_{p})}{(K_{p} + 2K_{f}) + \phi(K_{f}-K_{p})} \right].
\end{equation} 
Following Brinkman~\cite{brinkman_1952}, the effective  dynamic viscosity 
$\mu$ of a nanofluid may be modelled as: 
\begin{equation}
\mu = \mu_{f}(1-\phi)^{-2.5},
\end{equation}
where $\mu_f$ is the dynamic viscosity of the base fluid. All diffusion coefficients may then be computed using these expressions. The effective kinematic viscosity $\nu$ or the effective momentum diffusion coefficient of the nanofluid  may be computed as:
\begin{equation}
\nu = \frac{\mu}{\rho} = \frac{\mu_{f}(1-\phi)^{-2.5}}{(1-\phi){\rho}_{f} + \phi {\rho}_{p}}.
\end{equation}
Similarly the effective thermal diffusion coefficient $\kappa = K/(\rho c_V)$ of a nanofluid with spherical non-magnetic metallic particles may be computed using the expression:
\begin{eqnarray}
\kappa &=& \frac{K_f}{[(1-\phi)(\rho c_V)_f + \phi (\rho c_V)_p]} \nonumber\\ 
   &\times & \left[ \frac{(K_{p}+2K_{f}) - 2\phi(K_{f}-K_{p})}{(K_{p} + 2K_{f}) + \phi(K_{f}-K_{p})} \right].
\end{eqnarray}                                                                                      
Initially the fluid is at rest and the heat flux across the fluid layer is only due to conduction. The lower boundary of the nanofluid is maintained at temperature $T_b$, while the upper boundary is maintained at temperature $T_u = T_b - {\Delta T}$.  Here, $\beta  = (T_u - T_b)/d = {\Delta T}/d < 0$. The steady state temperature profile $T_s (z)$, density stratification $\rho_s (z)$ and the pressure field $P_s (z)$ across the nanofluid in conduction state~\cite{chandrasekhar_1961} are given by,
\begin{eqnarray}
T_s (z) &=& T_b + \beta z,\\
\rho_s (z) &=& \rho_0 \left[1 + \alpha \left(T_b - T_s (z)\right)\right],\\
P_s (z) &=& P_0 - \rho_0 g \left(z + \frac{1}{2}\alpha \beta z^2 \right),
\end{eqnarray}
where $T_b$ and $\rho_0$ are the reference values of the temperature and density fields at the bottom surface of the nanofluid. $P_0$ is a constant, which includes the magnetic pressure. As soon as $\beta$ is raised above a critical value $\beta_c$, the basic state of conduction becomes unstable and convective flow (${\bf v}\neq 0$) begins.
All the fields are perturbed and may be written as:
\begin{eqnarray}
\rho_s (z) \rightarrow \tilde{\rho}(x, y, z, t) &=& \rho_s (z) + \delta \rho (x, y, z, t),\\
T_s (z) \rightarrow T(x, y, z, t) &=& T_s (z) + \theta (x, y, z, t),\\
P_s (z) \rightarrow P(x, y, z, t) &=& P_s (z) + p (x, y, z, t),\\
{\bm{\mathrm{B}}}_0 \rightarrow {\bm{\mathrm{B}}} (x, y, z, t) &=& {\bm{\mathrm{B}}}_0 + \bm{\mathrm{b}} (x, y, z, t).
\end{eqnarray} 

All length scales are measured in units of the fluid thickness $d$ and time is measured in units of the free-fall time $\tau_{f} = 1/\sqrt{\alpha g \beta}$, where $g$ is the acceleration due to gravity. The fluid velocity $\bm{\mathrm{v}}(x,y,z,t)=(\mathrm{v_1}, \mathrm{v_2}, \mathrm{v_3})^{T}$, the perturbation in pressure due to flow $p(x,y,x,t)$, the convective temperature $\theta (x,y,z,t)$ and the induced magnetic field $\bm{\mathrm{b}}(x,y,z,t)$ are made dimensionless by $\sqrt{\alpha g \beta d^2}$, $\rho_0 \alpha g \beta d^2$, $ \beta d$ and  $\nu \sigma \mu_0 B_0$, respectively. The value of the effective magnetic Prandtl number $\mathrm{Pm}$ is of the order of $10^{-5}$ or less for terrestrial fluids including nanofluids. We therefore set the value of $\mathrm{Pm}$ equal to zero in this work. This makes the induced magnetic field 
$\bm{\mathrm{b}}$ a slaved variable. The RBM in nanofluids is then described by the following dimensionless equations:
\begin{eqnarray}
& D_t\bm{\mathrm{v}}=-\nabla p+\sqrt{\frac{\mathrm{Pr}}{\mathrm{Ra}}}\nabla^2\bm{\mathrm{v}}+\frac{\mathrm{Q}\mathrm{Pr}}{\mathrm{Ra}}\partial_z\bm{\mathrm{b}}+\theta\bm{\mathrm{e}}_3,\label{eq:mom-v}\\
&\nabla^2\bm{\mathrm{b}} = -\sqrt{\frac{\mathrm{Ra}}{\mathrm{Pr}}}\partial_z\bm{\mathrm{v}}, \label{eq:mag-b}\\
& {D_t \theta} = \sqrt{\frac{1}{\mathrm{Ra} \mathrm{Pr}}}\nabla^2\theta + {\mathrm{v}}_3, \label{eq:theta}\\
&\nabla\cdot\bm{\mathrm{v}} = \nabla\cdot\bm{\mathrm{b}}=0,\label{eq:cont}
\end{eqnarray}
where $ D_t \equiv \partial_t + (\bm{\mathrm{v}}\cdot\nabla)$ is the material derivative. In the above the dimensionless number Rayleigh number $\mathrm{Ra}$ is  defined as $\mathrm{Ra}$ $=$ $\frac{\alpha \beta gd^4}{\nu \kappa}$ $=$ $\frac{\rho_0 g \alpha \beta d}{(\rho_0 \nu \kappa /d^3)}$. It is a ratio of the buoyancy force per unit volume to the drag force per unit volume due to thermo-viscous effect. Other dimensionless external parameter for magnetoconvection is the Chandrasekhar's number $\mathrm{Q}$, which is a measure of the strength of the external magnetic field and it is a ratio of the Lorentz force per unit volume to the drag force due to magneto-viscous effect. It is defined as $\mathrm{Q}$ $=$ $\frac{\sigma B_0^2 d^2}{\rho_0\nu}$ $=$ $\frac{B_0^2/(\mu_0 d)}{(\rho_0 \nu \eta/d^3)}$.  It is also equal to square of the Hartmann number $\mathrm{H}=B_0 d \sqrt{\sigma/(\rho_0 \kappa)}$. It plays the role which Taylor number plays in RBC with Coriolis force~\cite{chandrasekhar_1961}.

Horizontal boundaries, located at $z=0$ and $z=1$, are considered to be thermally conducting and electrically nonconducting. Teflon or
ethylene-vinyl-acetate (EVA) composites may realize these conditions in an experiment~\cite{lee_etal_2008}. Horizontal boundaries made of good thermal conducting material and maintained at constant temperatures do not allow temperature fluctuations at the boundaries due to convective flow in the fluid. So the convective temperature field $\theta$ vanishes at the boundaries. Electrically  nonconducting surfaces do not allow current across the surface. Therefore, the vertical component of the current density $\bm{\mathrm{j}} = (\bm{ \nabla \times b})/\mu_0$ should also vanish at the horizontal boundaries. In addition, the induced magnetic field should be continuous at the boundaries.  The induced magnetic field $\bm{\mathrm{b}}^p$ in an electrically non-conducting plate of permeability $\mu_p$ must be  derivable from a potential~\cite{chandrasekhar_1961}. That is, 
\begin{eqnarray}
\bm{\mathrm{b}} &=& \bm{\mathrm{b}}^p ~~\mbox{at z = 0, 1}, \mbox{where}\\ \bm{\mathrm{b}}^p &=& \bm{\nabla} \Psi,~~\mbox{where} ~~ \nabla^2 \Psi = 0.
\end{eqnarray}
In the limit $\mathrm{Pm} \rightarrow 0$, as considered here, the boundary conditions of the induced magnetic field $\bm{\mathrm{b}}$ are dictated by Eq.~\ref{eq:mag-b}. This equation is satisfied when $\mathrm{b}_1$, 
$\mathrm{b}_2$ and $\partial_z \mathrm{b}_3$ vanish at the horizontal boundaries. This choice also ensures that $\mathrm{j}_3 = 0$ and $\bm{\nabla \cdot \mathrm{b}} = 0$ are automatically satisfied. The velocity boundary conditions on horizontal boundaries are assumed to be \textit{stress-free}, which are idealized boundary conditions. A good approximation for stress-free boundary conditions were realized in experiments by Goldstein and Graham~\cite{goldstein_graham_1969}. RBM at higher values of Chandrasekhar's number $\mathrm{Q}$ flows are not affected significantly due to stress-free boundary conditions. The relevant boundary conditions~\cite{chandrasekhar_1961} are then given as:
\begin{equation}
\partial_z \mathrm{v}_1 = \partial_z \mathrm{v}_2 = \mathrm{v}_3 =\mathrm{b}_1 = \mathrm{b}_2 = \partial_z \mathrm{b}_3 =\theta=0~\mbox{at}~z=0, 1. \label{eq:bound-v} 
\end{equation}

Let us denote magnetic field in the upper boundary as $\bm{\mathrm {b}}|_{z>1}$ and the same in the lower boundary as $\bm{\mathrm {b}}|_{z<0}$, respectively. Then 
\begin{equation}
\bm{\mathrm {b}}|_{z \ge 1} = \bm{\nabla} \Psi|_{z \ge 1}~~~~\mbox{and}~~~~ \bm{\mathrm {b}}|_{z \le 0} = \bm{\nabla} \Psi|_{z \le 0},
\end{equation}
where $\Psi|_{z \ge 1}$ and $\Psi|_{z \le 0}$ are scalar potentials in the regions $z > 1$ and $z < 0$, respectively.
The non-zero horizontal velocities of a nanofluid at the stress-free boundaries allow surface currents at the horizontal boundaries.  The continuity of the vertical component of the induced magnetic field at the horizontal boundaries ($z = 0, 1$) fixes the horizontal current.  

The effective thermal Prandtl number of the water based  nanofluids may be varied from $6.5$ to $4.0$, if the volume fraction of spherical copper nanoparticles are varied from $0.2\%$ to  $8.0\%$. The set of hydromagnetic system (Eqs.~\ref{eq:mom-v}-\ref{eq:bound-v}) is applicable to water based homogeneous nanofluids with non-magnetic metallic particles.  In the absence of nanoparticles ($\phi = 0$), the hydrodynamic system  represents magnetoconvection in geophysical fluids. The value of $\mathrm{Pr}$ for Earth's liquid outer core is approximated to be in a range from $0.1$ to $10$~\cite{olson_glatzmaier_1996}. Some liquid crystals have $Pr\sim 4.0$. These equations may also be useful in electrically conducting gases. The gases at high temperatures may conduct electricity as in a discharge tube. The range of $\mathrm{Pr}$ is chosen to cover different types of fluids. The critical Rayleigh number $\mathrm{Ra}_{c}(\mathrm{Q})$ for the onset of stationary magnetoconvection depends on the Chandrasekhar's number $\mathrm{Q}$. The critical wave number $k_{c}(\mathrm{Q})$, which is the wave number at the onset of convection, also depends on $\mathrm{Q}$. The expressions for $\mathrm{Ra}_{c}(\mathrm{Q})$ and $k_{c}(\mathrm{Q})$ are:
\begin{eqnarray}
& \mathrm{Ra}_{c}(\mathrm{Q}) = \frac{\pi^2 + k_{c}^2}{k_{c}^2}\big[ ( \pi^2 + k_{c}^2 )^{2} + \pi^{2}\mathrm{Q} \big],\label{eq:Ra}\\
& k_{c}(\mathrm{Q}) = \pi \sqrt{a_{+} + a_{-} - \frac{1}{2}},\label{eq:k}\\
 & a_{\pm} = \Bigg( \frac{1}{4} \Big[\frac{1}{2} + \frac{\mathrm{Q}}{\pi^{2}} \pm \big[ \big( \frac{1}{2} + \frac{\mathrm{Q}}{\pi^{2}} \big)^{2} - \frac{1}{4} \big]^{\frac{1}{2}}\Big] \Bigg)^{\frac{1}{3}}, \label{eq:a} 
\end{eqnarray}

The global heat flux across the fluid layer is defined by Nusselt number $\mathrm{Nu}$, which is a ratio of spatially averaged the total heat flux and the conductive heat flux. It is defined in terms of dimensionless vertical velocity $\mathrm{v}_3$  and convective temperature $\theta$ as:
\begin{equation}
\mathrm{Nu}(t) = 1 + \frac{\sqrt{\mathrm{Ra}~\mathrm{Pr}}}{V}\int_0^{\frac{2\pi}{k_{c}}}\int_0^{\frac{2\pi}{k_{c}}}\int_0^1{\mathrm{v}_3 \theta~ dxdydz},
\end{equation}
where $V=4\pi^2/k_c^2$ is the dimensionless volume of the simulation box. The Nusselt number, which is a function of time for unsteady magnetoconvection, depends on $\mathrm{Ra}$, $\mathrm{Pr}$ and $\mathrm{Q}$. Its time averaged value over a long period $T$ is denoted as 
$\langle{\mathrm{Nu}}\rangle = \frac{1}{T}\int_0^T{\mathrm{Nu} (t) dt}$. The quantity $\mathrm{v}_3 \theta$ represents the local heat flux in the vertical direction due to magnetoconvection.

\begin{table*}[ht]
  \begin{center}
\def~{\hphantom{0}}
  \begin{tabular}{c@{\hskip 0.35in}c@{\hskip 0.35in}c@{\hskip 0.35in}c@{\hskip 0.35in}c@{\hskip 0.35in}c@{\hskip 0.35in}c@{\hskip 0.35in}c@{\hskip 0.35in}c@{\hskip 0.35in}c}

\hline\hline\\
$\mathrm{Pr}$  & $\mathrm{Ra}$  & $\mathrm{Q}$ &$\mathrm{Nu}$ & $\epsilon^{u}$(est.)  &  $\epsilon^{u}$(comp.)  & $\epsilon^{\theta}$(est.)  & $\epsilon^{\theta}$(comp.)&$\mathrm{Nu}_{kin}$&$\mathrm{Nu}_{th}$\\
\hline\hline\\
$1.0$ & $3.04 \times 10^6$ & $0$ & $19.76$ & $0.0107$ & $0.0107$ & $0.0113$ & $0.0112$&$19.66$  &$19.56$  \\ \\
&   & $300$ & $20.44$ & $0.0111$ & $0.0110$ & $0.0117$ & $0.0115$& $20.18$ & $20.10$  \\ \\

& & $700$ & $20.59$ & $0.0112$ & $0.0111$ & $0.0118$ & $0.0116$& $20.35$ & $20.23$ \\ \\
$4.0$ & $5.0 \times 10^5$ & $0$ & $13.53$ & $0.0088$ & $0.0088$ & $0.0095$ & $0.0095$&$13.45$ & $13.40$  \\ \\
&  & $100$ & $13.63$ & $0.0089$ & $0.0089$ & $0.0096$ & $0.0094$&$13.58$  &$13.30$  \\ \\

  &   & $300$ & $11.80$ & $0.0076$ & $0.0075$ & $0.0083$ & $0.0081$& $11.61$ & $11.46$ \\ \\

& & $500$ & $10.85$ & $0.0069$ & $0.0068$ & $0.0076$ & $0.0074$&$10.62$&$10.46$\\ \\
$6.4$ & $5.0 \times 10^5$ & $0$ & $14.31$ & $0.0074$ & $0.0074$ & $0.0079$ & $0.0079$&$14.24$  & $14.17$ \\ \\
 &  & $100$ & $13.41$ & $0.0069$ & $0.0068$ & $0.0074$ & $0.0073$& $13.16$ & $13.06$  \\ \\

 &  & $300$ & $11.43$ & $0.0058$ & $0.0057$ & $0.0063$ & $0.0062$&$11.20$  &$11.09$  \\ \\

 &  & $500$ & $10.78$ & $0.0054$ & $0.0053$ & $0.0060$ & $0.0058$&$10.48$  &$10.37$  \\ \\
 
\hline\hline
\end{tabular}
\caption{ 
List of Nusselt number $\mathrm{Nu (Q)}$ for different values of Chandrasekhar's number $\mathrm{Q}$, kinetic energy dissipation rate $\epsilon^u$ and `thermal energy' dissipation rate $\epsilon^{\theta}$. These dissipation rates are compared with their estimated values using the formulas $\epsilon^{u} = (\mathrm{Nu}-1)/\sqrt{\mathrm{Ra}\mathrm{Pr}}$ and $\epsilon^{\theta} = \mathrm{Nu}/\sqrt{\mathrm{Ra}\mathrm{Pr}}$.}
\label{table1}
 \end{center}
 \end{table*}

\section{\label{sec:DNS}Direct Numerical Simulations}
\noindent
Direct numerical simulations are done using pseudo-spectral method. All fields are assumed to be periodic in the horizontal plane. The expansion of the relevant perturbations, consistent with the boundary conditions considered, are: 
\begin{eqnarray}
{\mathrm{v}_1} (x,y,z,t) &=& \sum_{l,m,n} U_{lmn}(t) e^{ik(lx+my)}\cos{(n\pi z)},\label{eq.U}\\
{\mathrm{v}_2} (x,y,z,t) &=& \sum_{l,m,n} V_{lmn}(t) e^{ik(lx+my)}\cos{(n\pi z)},\label{eq.V}\\
{\mathrm{v}_3} (x,y,z,t) &=& \sum_{l,m,n} W_{lmn}(t) e^{ik(lx+my)}\sin{(n\pi z)},\label{eq.W}\\
{\theta} (x,y,z,t) &=& \sum_{l,m,n} {\Theta}_{lmn}(t) e^{ik(lx+my)}\sin{(n\pi z)}, \label{eq.T}\\
{p} (x,y,z,t) &=& \sum_{l,m,n} P_{lmn}(t) e^{ik(lx+my)}\cos{(n\pi z)},\label{eq.P}
\end{eqnarray} 
where $U_{lmn}(t)$, $V_{lmn}(t)$, $W_{lmn}(t)$, $\Theta_{lmn}(t)$, and $P_{lmn}(t)$ are the Fourier amplitudes in the expansion of the fields $ v_{1}$, $v_{2}$, $v_{3}$, $\theta$, and $p$ respectively. The wave vector of perturbations in the horizontal plane is $\textbf{k} = lk\bm{\mathrm{e}}_1 + mk\bm{\mathrm{e}}_2$.  We have set $k=k_c(\mathrm{Q})$, where $k_c(\mathrm{Q})$ is the critical wave number for a given value of $\mathrm{Q}$.  The integers $l, m, n$  can take values compatible with continuity equation, which leads to the following condition.
\begin{equation}
ilk_c (\mathrm{Q}) U_{lmn} + imk_c(\mathrm{Q}) V_{lmn} + n\pi W_{lmn}=0.
\end{equation}
The expansions of the magnetic fields in the boundaries outside the nanofluids~\cite{basak_etal_2014} may be expressed as:
\begin{eqnarray}
\Psi|_{z \ge 1} &=& \sum_{l,m,n} (-1)^{n+1}\Psi_{lmn}(t)e^{ik(lx+my)}e^{\gamma (1-z)},\\
\Psi|_{z \le 0} &=& \sum_{l,m,n} \Psi_{lmn}(t)e^{ik(lx+my)}e^{\gamma z},
\end{eqnarray}
where $\gamma = k \sqrt{(l^2 + m^2)}$ and $\Psi_{lmn}(t)=\frac{n\pi W_{lmn}(t)}{\gamma (\gamma^2 +n^2 \pi^2)}$.
The spatial grid resolutions of the periodic box of size $L \times L \times 1$, where $ L = 2\pi/k_c(\mathrm{Q})$.  Spatial resolution of $128\times 128\times 128$ or $256\times 256\times 256$ grid points has been used for simulations presented here. As the Rayleigh number is raised above a critical value $\mathrm{Ra}_c (\mathrm{Q})$, while keeping the values of $\mathrm{Q}$ and $\mathrm{Pr}$ fixed,  stationary magnetoconvection begins~\cite{chandrasekhar_1961}. We define the distance from criticality by a parameter $\epsilon$ $=$ $\frac{\mathrm{Ra}-\mathrm{Ra}_c (\mathrm{Q})}{\mathrm{Ra}_c (\mathrm{Q})}$.  As $\mathrm{Q}$ is raised keeping $\mathrm{Ra}$ and $\mathrm{Pr}$ fixed, the parameter $\epsilon$ becomes smaller and consequently the fluctuations are reduced. The fluctuations are more at lower values of $\mathrm{Q}$, if  the value of  $\mathrm{Ra}$ is sufficiently raised. As a results the spatial resolution required is less, if $\mathrm{Q}$ is raised to a higher value with $\mathrm{Ra}$ and $\mathrm{Pr}$ fixed. The spatial resolutions used here are sufficient to describe the magnetoconvective flow for the parameter values considered. We have compared our results for $\mathrm{Q} = 0$ with those obtained by Veronis~\cite{veronis_jfm_1966}, Moore \& Weiss~\cite{moore&weiss_jfm_1973} and Thual~\cite{thual_jfm_1992}, who used the identical boundary conditions. The two sets of grid resolutions used here keep the minimum value of the global Kolmogorov dissipative scale always more than $2$. We have also reproduced the results reported in the earlier works on magnetoconvective instability~\cite{basak_etal_2014} as well as on RBC~\cite{hp_kk_jkb_pre_2014}. Of course, for much lower values of $\mathrm{Pr}$ ($ < 0.1$) and much higher values of $\mathrm{Ra}$ would require better spatial grid resolutions. We have listed in Table~\ref{table1} values of Nusselt number $\mathrm{Nu (Q)}$, the average dissipative rates for the kinetic energy $\epsilon^u = \langle\nu\int_0^{\frac{2\pi}{k_{c}}}\int_0^{\frac{2\pi}{k_{c}}}\int_0^1\{\partial_i v_j(x,y,z,t)\}^2 dx dy dz\rangle$ and `thermal energy' $\epsilon^{\theta} = \langle\kappa\int_0^{\frac{2\pi}{k_{c}}}\int_0^{\frac{2\pi}{k_{c}}}\int_0^1\{\partial_i \theta(x,y,z,t)\}^2 dx dy dz\rangle$ for different values of Chandrasekhar's number $\mathrm{Q}$. The computed values of the dissipation rates $\epsilon^u$ and $\epsilon^{\theta}$ are compared with their estimated values using the formulas: $\epsilon^u = (\mathrm{Nu (Q)}-1)/\sqrt{\mathrm{Ra Pr}}$ and $\epsilon^{\theta} = \mathrm{Nu (Q)}/\sqrt{\mathrm{Ra Pr}}$ in Table~\ref{table1}. They are in good agreement.  Rayleigh-B\'{e}nard convection was investigated numerically in a cubic box with no-slip velocity boundary conditions on all walls by Xu etal~\cite{Xu_etal_ijhmt_2019}. Their velocity boundary conditions and simulation box size were different than what we have considered here. The values for the Nusselt number in a fluid with $\mathrm{Pr} = 7.0$ are $8.49$ and $11.12$ for $\mathrm{Ra} = 10^6$ and $3\times 10^6$, respectively. Our values for Nusselt number at are almost double ($17.32$ for $\mathrm{Ra} = 10^6$ and $23.54$ for $\mathrm{Ra} = 3\times 10^6$). The definitions of the dissipation rate for kinetic energy also differs by a numerical factor of $1/2$. 

We record the values of all relevant fields at all spatial grid points at an regular interval of every two units of dimensionless time. We have computed minimum number of 300 frames for each set of parameter values reported here. 

\begin{figure}[ht]
\begin{center}
\includegraphics[height=!, width=9.0 cm]{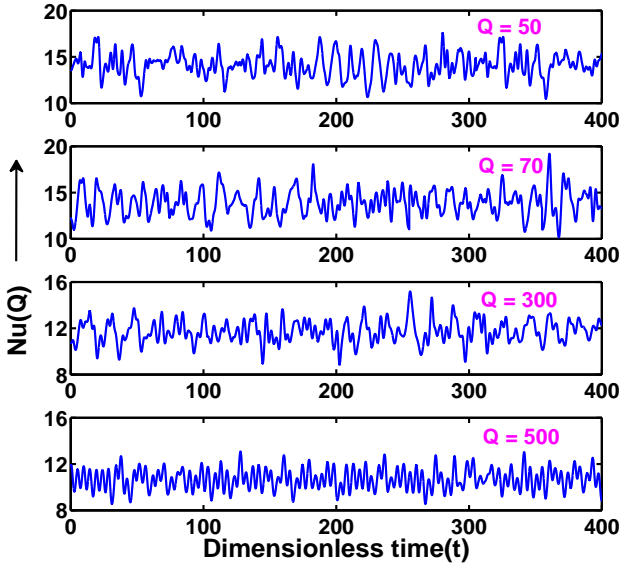} 
\caption{\label{fig:Nu_time} 
(Color online) Variation of Nusselt number $\mathrm{Nu(Q)}$ with dimensionless time $t$ [blue (black) curves] for Rayleigh number $\mathrm{Ra} = 5.0\times 10^5$ and Prandtl number $\mathrm{Pr=4.0}$ for different values of Chandrasekhar number $\mathrm{Q}$.}
\end{center}
\end{figure}

\begin{figure}[htp!]
\includegraphics[height=!, width=9.0 cm]{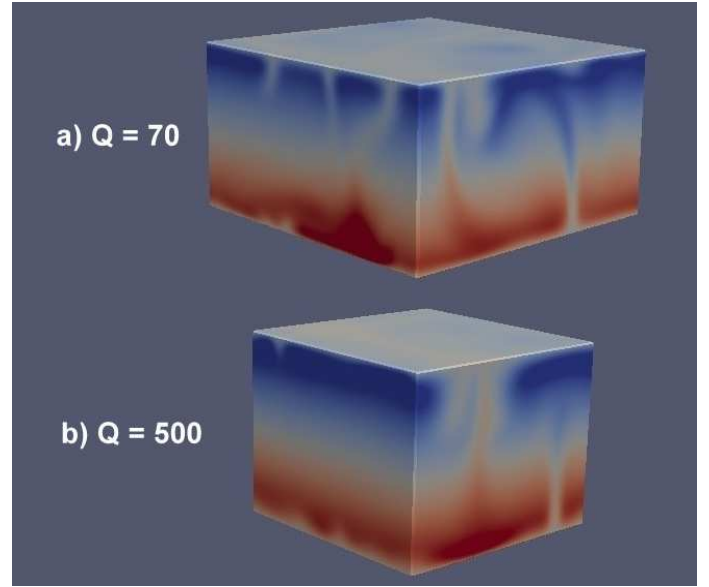} 
\caption{(Color online) Three-dimensional temperature isosurfaces computed from DNS for  $\mathrm{Ra} = 5.0 \times 10^5$ and $\mathrm{Pr} = 4.0 $ for  (a) $\mathrm{Q} = 70$ and  (b) $\mathrm{Q} = 500$. Red (grey) and blue (black) colors stand for hotter and cooler fluids, respectively.}
\label{thermal_plumes}a
\end{figure}
\begin{figure}[htp!]
\begin{center}
\includegraphics[height=!, width=9.0 cm]{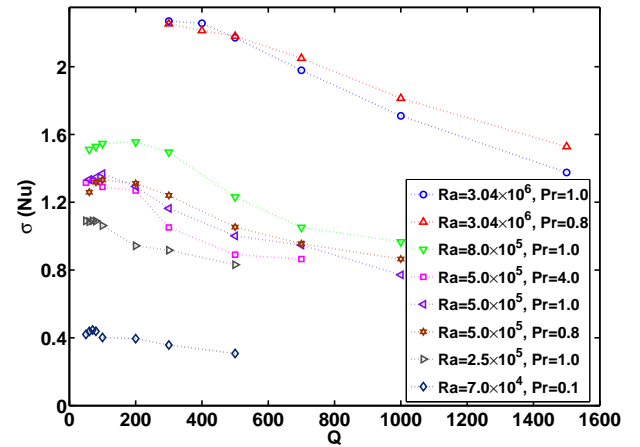} 
\caption{\label{fig:sigma_Nu_fluctuations} 
(Color online) Standard deviation $\sigma (\mathrm{Nu})$ of the Nusselt number 
$\mathrm{Nu}$ for different values of $\mathrm{Ra}$, $\mathrm{Pr}$ and $\mathrm{Q}$.}
\end{center}
\end{figure}

\begin{figure}[ht]
\begin{center}
\includegraphics[height=!, width=9.0 cm]{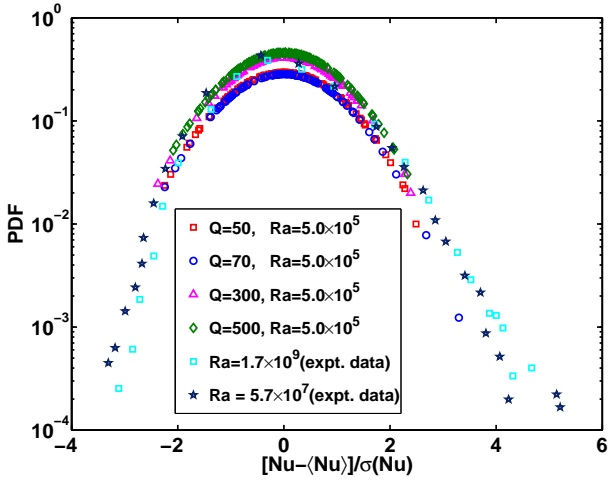} 
\caption{\label{fig:PDF_Nu_fluctuations} 
(Color online) Probability density function (PDF) of the fluctuations in Nusselt number $\mathrm{Nu}$ at $\mathrm{Ra} = 5.0\times 10^5$ and $\mathrm{Pr=4.0}$ for different values of $\mathrm{Q}$. Red (gray) squares, blue (black) circles, magenta (dark gray) triangles and green (gray) diamonds  are for $\mathrm{Q} = 50$, $70$, $300$ and $500$, respectively. Blue (black) stars and cyan (light gray) squares are experimental data points (Ref.~\cite{aumaitre_fauve_epl_2003}) in absence of the external magnetic field ($\mathrm{Q} =0$) for $\mathrm{Ra} = 5.7 \times 10^7$ and $1.7 \times 10^9$, respectively.}
\end{center}
\end{figure}

\begin{figure}[ht]
\begin{center}
\includegraphics[height=!, width=9.0 cm]{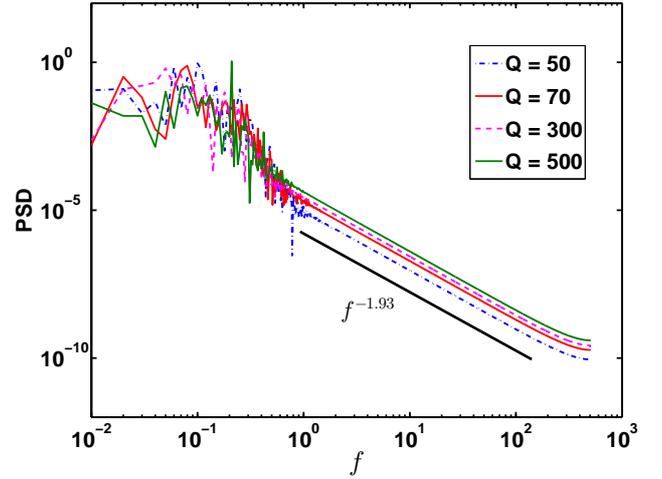} 
\caption{\label{fig:Nu_PSD} 
(Color online) Frequency power spectral density (PSD) of the Nusselt number ($\mathrm{Nu}$) for $\mathrm{Ra} = 5.0\times 10^5$ and $\mathrm{Pr=4.0}$ at different values of $\mathrm{Q}$.}
\end{center}
\end{figure}

\section{\label{sec:Result}Results and Discussions}
\noindent
 As Rayleigh number $\mathrm{Ra}$ is raised to a sufficiently high value for fixed values of Chandrasekhar number $\mathrm{Q}$ and Prandtl number 
$\mathrm{Pr}$, the magnetoconvection becomes unsteady. Fig~\ref{fig:Nu_time} shows the variation of Nusselt number for chaotic magnetoconvective flow with dimensionless time for $\mathrm{Ra} = 5.0\times 10^5$ and $\mathrm{Pr=4.0}$ for different values of $\mathrm{Q}$. The temporal evolution of Nusselt number for $\mathrm{Q} = 70, 300, 500$ show that the time averaged mean value $\mathrm{Nu}(\mathrm{Q})$ decreases with increase in $\mathrm{Q}$. It confirms that the magnetic field suppresses the heat flux  of in RBM at relatively larger values of $\mathrm{Q}$. Figure~\ref{thermal_plumes} shows typical three-dimensional isosurfaces computed at a given instant from the DNS for $\mathrm{Ra}= 5\times 10^5$ and $\mathrm{Pr}=4.0$ for two different values of $\mathrm{Q}$. More thermal plumes are generated for $\mathrm{Q} = 70$ than for $\mathrm{Q} = 500$. The generation of more thermal plumes leads to enhancement of the relative Nusselt number in a range of lower values of $\mathrm{Q}$ for fixed values of $\mathrm{Ra}$ and $\mathrm{Pr}$.

\begin{figure*}[htp]
\begin{center}
\includegraphics[height=!, width=17.0 cm]{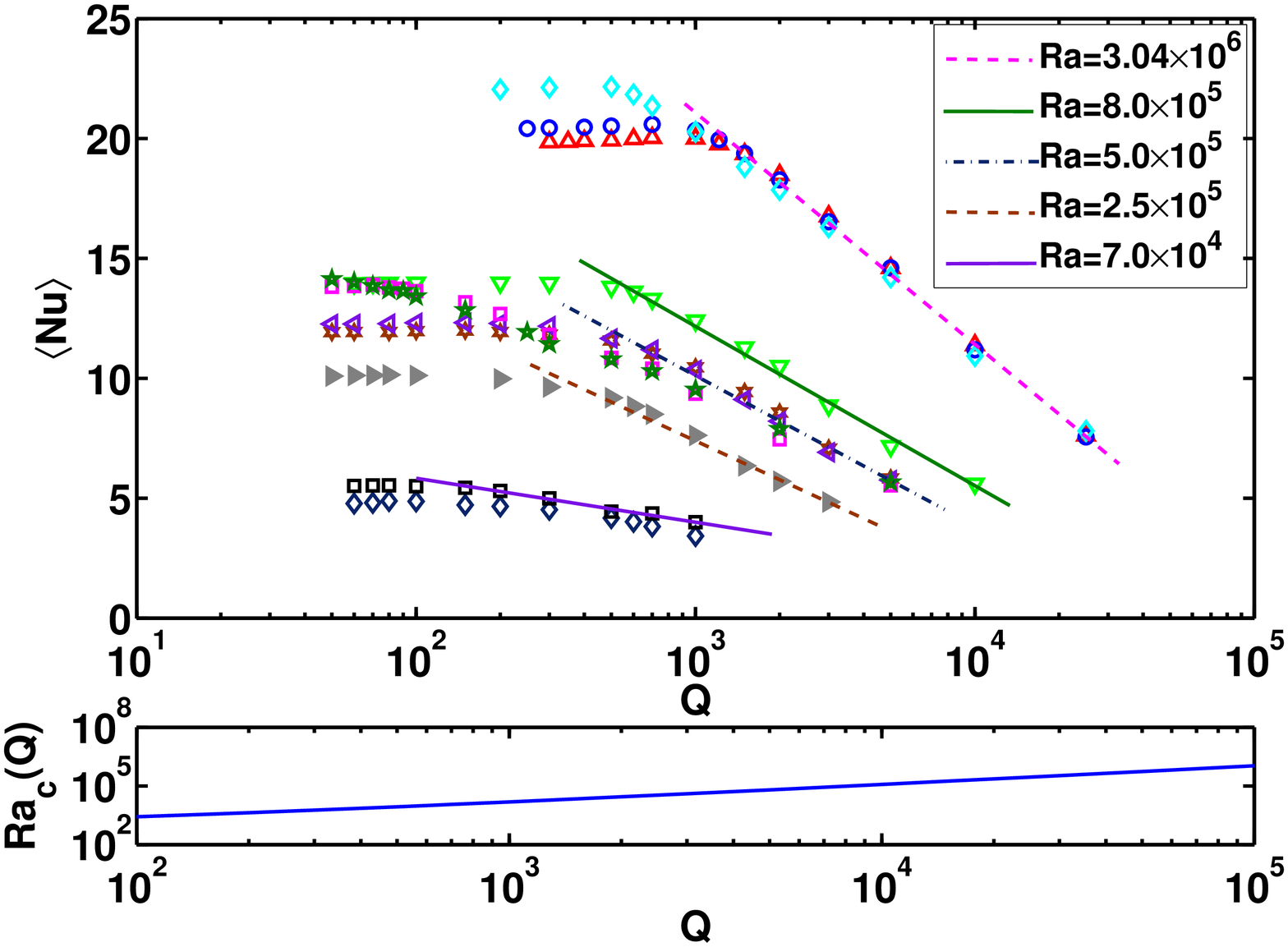} 
\caption{\label{fig:Nu_Q} 
(Color online) The variation of time averaged Nusselt number $\langle \mathrm{Nu}\rangle$ with Chandrasekhar's number $\mathrm{Q}$ for different values of $\mathrm{Ra}$ and $\mathrm{Pr}$ is shown in the upper viewgraph. The data set at the top is for $\mathrm{Ra} = 3.04 \times 10^6$. Data points shown by cyan (light gray) diamonds, blue (black) circles and red (dark gray) triangles are computed  for $\mathrm{Pr} = 2.0$, $1.0$ and $0.8$, respectively. The second set of computed data points from the top [green (gray) inverted triangles] are for $\mathrm{Ra} = 8.0 \times 10^5$ and $\mathrm{Pr} = 1.0$. The third data set from the top is for $\mathrm{Ra} = 5.0 \times 10^5$. Here, brown (dark gray) stars, violet (gray) left pointing triangles, magenta (light gray) squares  and  green (gray) stars are data points computed for $\mathrm{Pr} = 0.8$, $1.0$, $4.0$ and $6.4$ respectively.The fourth data set from the top [azure (gray) right pointing triangles] is for $\mathrm{Ra} = 2.5 \times 10^5$ and $\mathrm{Pr} = 1.0$. The set of data points at the bottom  are for $ \mathrm{Ra} = 7.0 \times 10^4$. Here blue (black) diamonds and black (black) squares are data points computed for $\mathrm{Pr}=0.1 $ and $\mathrm{Pr}=0.2 $, respectively. The lower viewgraph shows the plot of the threshold $\mathrm{Ra}_c (\mathrm{Q})$ with $\mathrm{Q}$.}
\end{center}
\end{figure*}
\begin{figure*}
\begin{center}
\includegraphics[height=!, width=17.0 cm]{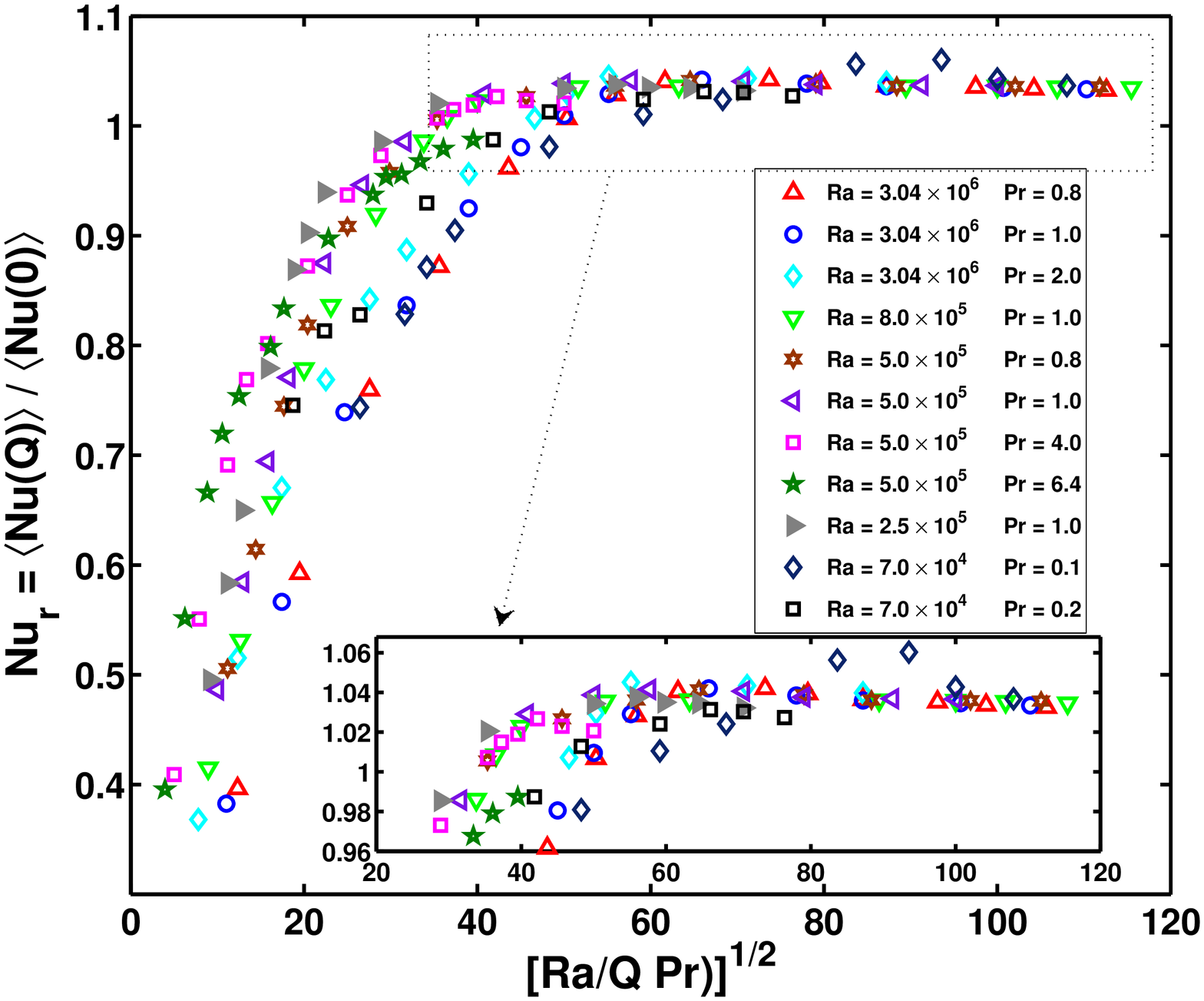} 
\caption{\label{fig:Nu_enhancement} 
(Color online) Variation of the reduced Nusselt number $\mathrm{Nu_r}$ $=$ $\langle \mathrm{Nu}(\mathrm{Q})\rangle/\langle \mathrm{Nu}(0)\rangle$ with the dimensionless quantity $\sqrt{\mathrm{Ra/(Q  Pr)}}$ for several sets of $\mathrm{Ra}$ and $\mathrm{Pr}$: (i) $\mathrm{Pr}=0.1 $ with $ \mathrm{Ra} = 7.0 \times 10^4$ [blue (black) diamonds], (ii) $\mathrm{Pr}=0.2 $ with $ \mathrm{Ra} = 7.0 \times 10^4$ [black (black) squares],
(iii) $\mathrm{Pr} = 0.8 $ and $\mathrm{Ra} = 3.04 \times 10^6$ [red (gray) triangles] and $5.0 \times 10^5$ [brown (dark gray) stars], (iv) $\mathrm{Pr}=1.0 $ with $ \mathrm{Ra} = 3.04 \times 10^6$ [blue (black) circles], $ 8.0 \times 10^5$ [green (light gray)  inverted triangles], $ 5.0 \times 10^5$ [violet (gray) left pointing triangles] and $2.5 \times 10^5 $ [azure (gray) right pointing triangles], (v) $\mathrm{Pr}=2.0$ with $\mathrm{Ra} = 3.04 \times 10^6 $ [cyan (light gray) diamonds], (vi) $\mathrm{Pr}=4.0$ with $\mathrm{Ra} = 5.0 \times 10^5$  [magenta (light gray) squares] and (vii) $\mathrm{Pr}=6.4$ with $\mathrm{Ra} = 5.0 \times 10^5$ [green (gray) stars]. Inset shows the enlarged view of the plot for $\mathrm{Nu_r} \approx 1.0$.}
\end{center}
\end{figure*}

\begin{figure}[ht]
\begin{center}
\includegraphics[height=!, width=9.0 cm]{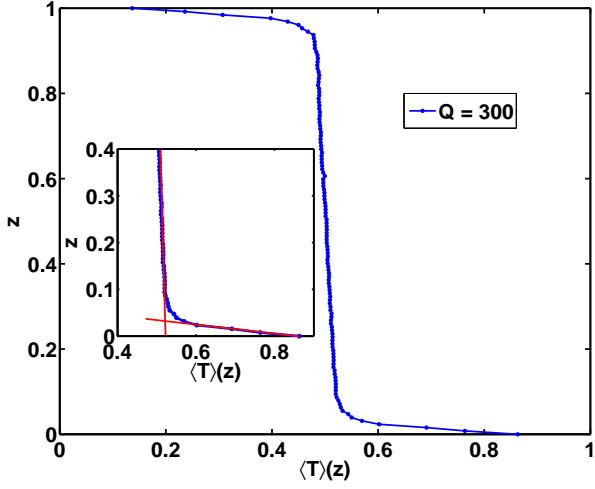} 
\caption{\label{fig:temperature} 
(Color online) Variations of horizontally averaged temperature field $\langle \mathrm{T}\rangle (z)$ fields with the vertical coordinate are plotted for $\mathrm{Q} = 300$ [blue (black) circle], for $\mathrm{Pr} = 1.0$ and $\mathrm{Ra} = 3.04 \times 10^6$. Red (gray) straight lines show the variations of temperature
in the central part and near the lower boundary.}
\end{center}
\end{figure}
\begin{figure}[ht]
\begin{center}
\includegraphics[height=!, width=9.0 cm]{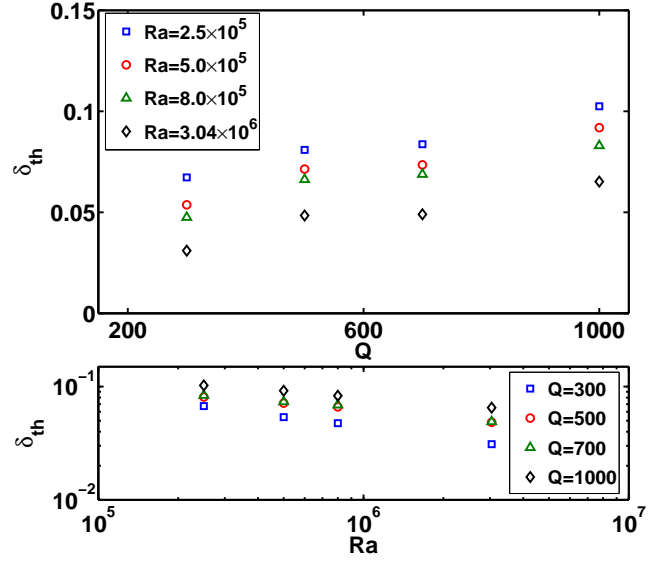} 
\caption{\label{fig:boundary_layer} 
(Color online) Thermal boundary layer: Plot of the thickness of thermal boundary layer 
($\delta_{\mathrm{th}}$) computed for $\mathrm{Pr} = 1.0$ and $\mathrm{Ra}$ and $\mathrm{Q}$. The upper viewgraph shows the variation of $\delta_{\mathrm{th}}$ with $\mathrm{Q}$ for different values of $\mathrm{Ra}$. The lower viewgraph shows variation of $\delta_{\mathrm{th}}$ with $\mathrm{Ra}$ (on log-log scale) for different values of $\mathrm{Q}$. }
\end{center}
\end{figure}

 \begin{figure}[htp!]
\includegraphics[height=!, width=9.0 cm]{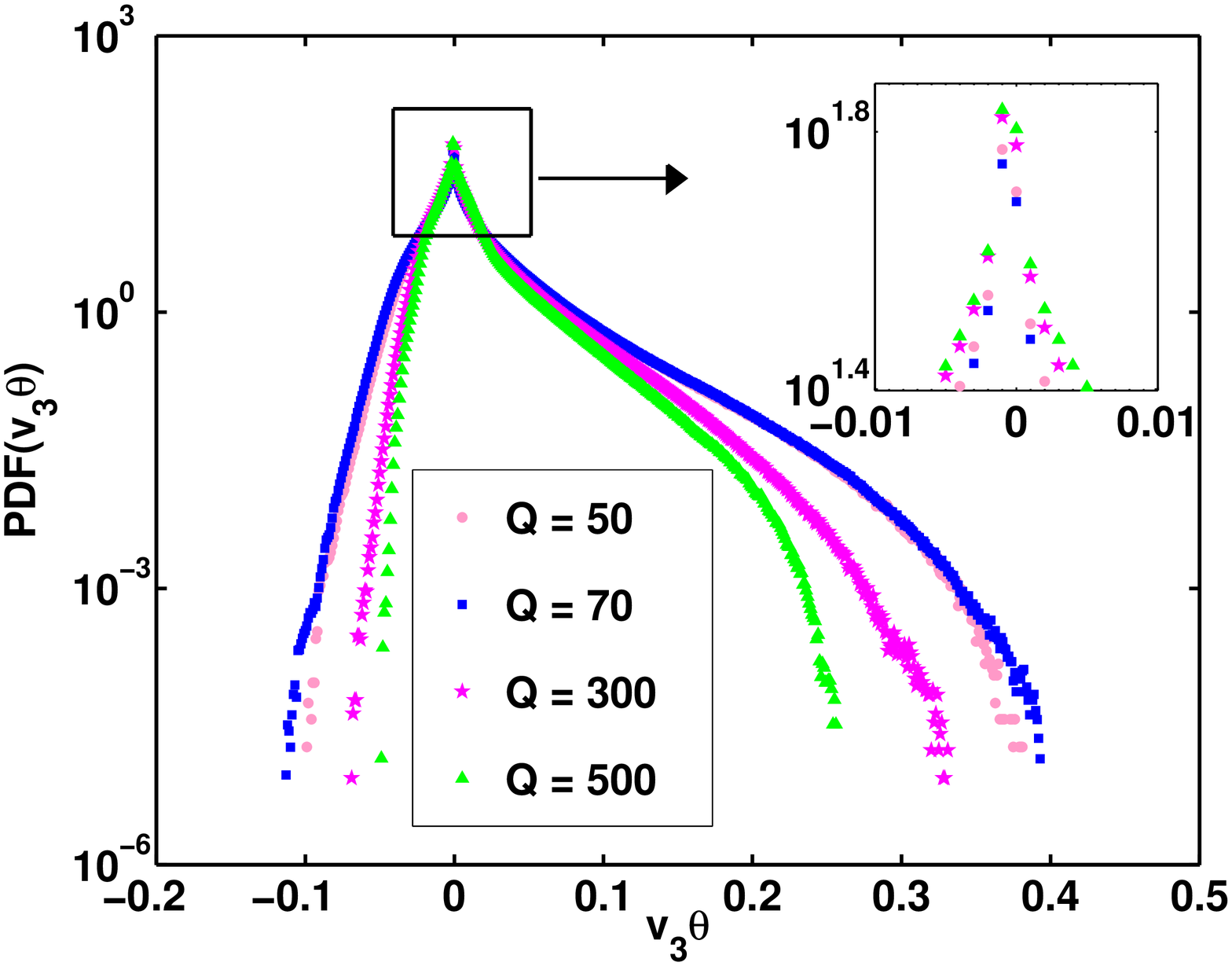} 
\caption{\label{fig:Pr_4.0} 
(Color online) The probability distribution functions (PDFs) of vertical heat flux for $\mathrm{Pr} = 4.0 $ computed for $\mathrm{Q}= 50$ [pink (light gray) circles], $70$ [blue (black) squares], $300$ [magenta (gray) stars], and $500$ [green (gray) triangles] with $\mathrm{Ra} = 5.0 \times 10^{5}$. Inset shows the PDFs near their maxima.}
\end{figure}

The temporal fluctuation of Nusselt number is considerable for lower values of $\mathrm{Q}$ (see Fig.~\ref{fig:Nu_time}). Figure~\ref{fig:sigma_Nu_fluctuations} displays the standard deviation in the temporal signals of Nusselt number for different values of $\mathrm{Ra}$, $\mathrm{Pr}$ and $\mathrm{Q}$. Fluctuations are larger at higher values $\mathrm{Ra}$. Fluids with lower values of effective thermal Prandtl   number $\mathrm{Pr}$ show relatively larger fluctuations. The fluctuations are suppressed at higher values of the applied magnetic field, if all other parameters are kept fixed. For higher values of $\mathrm{Q}$, the distance from the onset of magnetoconvection is smaller and, consequently, the fluctuating part of the Nusselt number is reduced. 

Figure~\ref{fig:PDF_Nu_fluctuations} shows the probability distribution function (PDF) of the fluctuating parts of the Nusselt number around the mean for different values of $\mathrm{Q}$. Probability distribution function is close to normal distribution but with slightly asymmetric tails.  For effective value of $\mathrm{Pr} = 4$, the height of PDF is the lowest for $\mathrm{Q} = 70$ but tails are longer. As $\mathrm{Q}$ is increased or decreased, the height of PDF goes up and tails become shorter. We have  also compared the computed PDFs with  the  experimental results of Aumaitre and Fauve~\cite{aumaitre_fauve_epl_2003} on Rayleigh-B\'{e}nard convection (RBC) in water and mercury in the absence of external magnetic field.
The data points shown by navy blue (black) stars and cyan (light gray) squares  are adopted from their experiments at much higher values of $\mathrm{Ra}$.  PDFs obtained for RBM are in good agreement with those observed in RBC in water and mercury.  The slight difference is due to small Rayleigh number used for simulations. The computed curves are smoother due to large number of data points. 

Figure~\ref{fig:Nu_PSD} shows the power spectrum density (PSD) in the frequency space of the Nusselt number for $\mathrm{Ra} = 5.0 \times 10^5$ and $\mathrm{Pr} = 4.0$ and for different values of $\mathrm{Q}$. The PSD shows noisy behaviour at lower frequencies. However, the Nusselt number is found to show scaling behaviour at higher frequencies. The PSD of $\mathrm{Nu}$ scales with frequency $f$ approximately as  $f^{-2}$. The best fit to curves obtained from simulations gives the value of the exponent as $-1.93\pm 0.02$. The value of the exponent is quite close to one observed in experiments on turbulent RBC~\cite{aumaitre_fauve_epl_2003} as well as in numerical simulations of turbulent RBC with rotation~\cite{pharasi_etal_pre_2014}. 
 
The upper viewgraph in Fig.~\ref{fig:Nu_Q} displays the variation of time averaged Nusselt number $\langle \mathrm{{Nu}(Q)}\rangle$ with Chandrasekhar number $\mathrm{Q}$ for several values of $\mathrm{Ra}$ and $\mathrm{Pr}$ on a semi-log scale. The magenta (very light gray), green (light gray), blue (black), brown (dark gray) and violet (gray) curves are the best fit to the data points obtained from simulations  for different curves. Cyan (light gray) diamonds, blue (black) circles and red (gray) triangles at the top are the  data points computed for $\mathrm{Pr} = 2.0 $, $1.0$ and $0.8$, respectively. The green (gray) inverted triangles represent the data points computed for $\mathrm{Pr} = 1.0$ in the second data set from top ($\mathrm{Ra}= 8.0 \times 10^5$). Brown (dark gray) stars, violet (gray) left pointing triangles, magenta (light gray) squares and green (gray) stars in the third data set ($\mathrm{Ra}=5.0 \times 10^5$) are computed for $\mathrm{Pr} = 0.8$, $1.0$, $4.0$ and $6.4$, respectively. Data points, shown as azure (gray) right pointing triangles in the fourth data set from the top ($\mathrm{Ra} = 2.5 \times 10^5$), are for $\mathrm{Pr} = 1.0$. Data points, shown as blue (black) diamonds and black (black) squares  in the data set at the bottom are for $\mathrm{Pr}=0.1$ and $\mathrm{Pr}=0.2$, respectively. The lower viewgraph in Figure~\ref{fig:Nu_Q} shows the plot of threshold $\mathrm{Ra}_c (\mathrm{Q})$ for stationary magnetoconvection with $\mathrm{Q}$, as obtained by Chandrasekhar~\cite{chandrasekhar_1961} for stress-free velocity boundary conditions. The time averaged value of the Nusselt number $\langle\mathrm{Nu(Q}\rangle$, for fixed values of $\mathrm{Ra}$ and $\mathrm{Pr}$, first increases very slowly with $\mathrm{Q}$, reaches a maximum  and then starts decreasing quickly with $\mathrm{Q}$. The tendency of slight enhancement of heat flux  was not observed for $\mathrm{Pr} = 6.4$. It has some similarity with enhancement of thermal flux at low rotation rates in rotating RBC. Fig.~\ref{fig:Nu_Q} also shows that decrease of $\langle \mathrm{Nu(Q)}\rangle$ with $\mathrm{Q}$ is logarithmic for higher values of  $\mathrm{Q}$ [see the magenta (light gray), green (gray), blue (black) and brown (dark gray) lines].  For given values of $\mathrm{Q}$ and $\mathrm{Pr}$, the mean Nusselt number is higher for larger values of $\mathrm{Ra}$. The effect of $\mathrm{Pr}$ is clearly visible only for  $\mathrm{Q} < \mathrm{Q}_c$ (where  $\mathrm{Q} = \mathrm{Q}_c$ denotes the critical value of Chandrasekhar number, above which the logarithmic behaviour starts to set in), if $\mathrm{Ra}$ is kept fixed. For $\mathrm{Q} < \mathrm{Q}_c$, $\langle \mathrm{Nu} \rangle$ increases with $\mathrm{Pr}$. There is no significant change in $\langle \mathrm{Nu} \rangle$ for $\mathrm{Q} > \mathrm{Q}_c$ with $\mathrm{Pr}$, if $\mathrm{Ra}$ is kept fixed. However, $\mathrm{Nu}_c$ (the critical value of time averaged Nusselt number at $\mathrm{Q}_c$) increases but $\mathrm{Q}_c$ decreases with $\mathrm{Pr}$ for a fixed value of $\mathrm{Ra}$. The values of $\mathrm{Nu}_c$ and $\mathrm{Q}_c$ both increases with $\mathrm{Ra}$. For $\mathrm{Pr}=1.0$, we observe that $\mathrm{Nu}_c$ and $\mathrm{Q}_c$ vary with $\mathrm{Ra}$ as ${\mathrm{Ra}}^{0.28\pm 0.01}$ and   ${\mathrm{Ra}}^{0.65\pm 0.03}$, respectively. For $\mathrm{Q} > \mathrm{Q}_c$, the Lorentz force starts playing dominant role on the heatflux across the fluid layer. On the other hand, the role of Lorentz force is less significant for $\mathrm{Q} < \mathrm{Q}_c$. 

The slope of the $\langle \mathrm{Nu(Q)}\rangle-\mathrm{Q}$ curves for $\mathrm{Q}$ depends mainly on $\mathrm{Ra}$. The time averaged Nusselt number for $\mathrm{Q} > \mathrm{Q}_c$ may therefore be expressed  as:
\begin{equation}
\langle \mathrm{Nu(Q)}\rangle = C_1(\mathrm{Ra},  \mathrm{Pr}) - C_2 (\mathrm{Ra},  \mathrm{Pr}) \ln{ \mathrm{Q}},
\end{equation}
where $C_1$ and $C_2$ are functions of $\mathrm{Ra}$ and $\mathrm{Pr}$. 

We now define a reduced (or normalized) Nusselt number $\mathrm{Nu_r}$ $=$ $\langle \mathrm{Nu(Q)}\rangle /\langle \mathrm{Nu(0)}\rangle$ as a ratio of the Nusselt number in the presence of an external magnetic field ($\mathrm{Q}\neq 0$) and the Nusselt number $\mathrm{Nu(0)}$ in the absence of any external magnetic field ($\mathrm{Q} = 0$). The dimensionless parameter $\mathrm{Ra/(Q Pr)}$ is a ratio of the buoyancy force per unit volume ($\alpha \beta g d^4 \rho_0$) and the Lorentz force per unit volume ($\sigma \mathrm{B}_0^2 d^2 \nu $).  If the vertical magnetic field always suppressed the transport of heat across the fluid layer~\cite{chandrasekhar_1961,cioni_etal_2000,aurnou_olson_2001,burr_mueller_2001}, the value of $\mathrm{Nu}_r$ should always be less than unity and it should approach asymptotically to unity as the parameter $\sqrt{\mathrm{Ra/(Q Pr)}}$ is raised to a much larger value. Fig.~\ref{fig:Nu_enhancement} shows the variation of  $\mathrm{Nu_r}$ with $\sqrt{\mathrm{Ra}/\mathrm{(Q Pr)}}$ for different values of $\mathrm{Pr}$.  $\mathrm{Nu_r}$ increases sharply from a small value ($\ll 1$) and attains a value  slightly bigger than unity, as $\sqrt{\mathrm{Ra}/\mathrm{(Q Pr)}}$ is raised slowly. With further increase in  $\sqrt{\mathrm{Ra}/\mathrm{(Q Pr)}}$, the value of $\mathrm{Nu_r}$ starts decreasing slowly and tends to approach unity (see the plots for $\mathrm{Pr} = 0.8, 1.0, 4.0$) slowly. The inset in Fig.~\ref{fig:Nu_enhancement} shows an enlarged view of the curve showing $\mathrm{Nu_r}$ more than unity. The maximum enhancement of thermal flux is observed for $0.1 \le \mathrm{Pr} \le 4.0$. The value of the parameter $\sqrt{\mathrm{Ra}/\mathrm{(Q Pr)}}$, where $\mathrm{Nu_r}$ reaches its maximum, depends on $\mathrm{Ra}$ and $\mathrm{Pr}$. We do not observe enhancement of thermal flux for $\mathrm{Pr} = 6.4$. For lower values of $\mathrm{Q}$  and for a range  of $\mathrm{Pr}$, the enhancement of thermal flux is observed in the unsteady  magnetoconvection. This behaviour has similarity with enhancement of thermal flux observed in rotating RBC at lower values Rossby number (higher rotation rates)~\cite{liu_ecke_prl_1997,stevens_etal_2009,zhong_etal_prl_2009,weiss_etal_prl_2010,stevens_etal_njp_2010,wei_etal_prl_2015}.
However, the amount of enhancement observed in the case of magnetoconvection is less compared to that observed in rotating RBC. In addition, the enhancement of thermal flux in magnetoconvection is not observed at larger values of $\mathrm{Pr}$ in RBM. This may be due to efficient generation of thermal plumes in rotating RBC at relatively higher values of $\mathrm{Pr}$~\cite{stevens_etal_njp_2010} in rotating RBC.  

Thin boundary layers are also characteristics of a turbulent flow~\cite{Zhou_Xia_prl_2010,Zhou_Xia_jfm_2013}. The thickness of thermal boundary layer $\delta_{th}$ in turbulent RBC is known to scale with $\mathrm{Ra}$ as $\delta_{th} \sim \mathrm{Ra}^{-\gamma}$. The exponent $\gamma$ is found to lie between $0.2$ and $0.3$~\cite{Zhou_Xia_jfm_2013}.  We also investigated the role of magnetic field on the thickness of thermal boundary layer. To compute the thickness of the boundary layer ($\delta_{th}$), we first spatially averaged the total temperature field $T(x,y,z,t)$ in horizontal plane for each frame of computed data points. This led to a temperature field, which is a function of the vertical coordinate $z$ and dimensionless time $t$. A time average of a large number of frames (300 frames or more) yielded a temperature field $\langle T \rangle (z)$, which depends only on the vertical coordinate $z$. One such case for $\mathrm{Ra} = 3.04 \times 10^6$, $\mathrm{Pr} = 1.0$ and $\mathrm{Q}=300$ is shown in Fig.~\ref{fig:temperature}. It clearly shows a sharp drop in the temperature field in a thin layer of the fluid near both the boundaries. The temperature drop in the central part of the simulation cell is very small. We draw two straight lines: one drawn through the almost vertical part and another drawn through the part where the temperature drop is sharp (see Fig.~\ref{fig:temperature}). The estimated thermal boundary layer $\delta_{th}$ is the vertical distance of the point of intersection from the nearest boundary. The upper viewgraph of Fig.~\ref{fig:boundary_layer} shows the variation of the thickness of thermal boundary layer with $\mathrm{Q}$ for $\mathrm{Pr} = 1.0$ and different values of $\mathrm{Ra}$. The thickness $\delta_{th}$ increases with $\mathrm{Q}$ for a fixed value of $\mathrm{Ra}$. It is expected as the increase in $\mathrm{Q}$ brings down the distance from criticality $\epsilon$. The lower viewgraph shows the variation of $\delta_{th}$ with $\mathrm{Ra}$ for different values of $\mathrm{Q}$ on log-log scale. The boundary layer thickness decreases with increase in $\mathrm{Ra}$ for a fixed value of $\mathrm{Q}$. The boundary layer thickness shows scaling behavior with $\mathrm{Ra}$: $\delta_{th} \sim \mathrm{Ra}^{-\gamma}$, where the exponent $\gamma (\mathrm{Q})$ now depends on the Chandrasekhar's number $\mathrm{Q}$. The value of $\gamma$ is found to be $0.30 \pm 0.01$ for $\mathrm{Q} = 3\times 10^2$ and $0.18 \pm 0.01$ for $\mathrm{Q} = 10^3$. The value of $\gamma$ is in excellent agreement with the experimental observation of Zhou and Xia~\cite{Zhou_Xia_jfm_2013} for lower value of $\mathrm{Q}$. 

We have also computed the probability distribution functions (PDFs) of the local heat fluxes in the vertical direction to investigate the role of the external magnetic field on PDFs. For this, the values of the vertical velocity $\mathrm{v}_3$ and convective temperature $\theta$ are recorded at all spatial grid points at regular interval for a long time.  A probability distribution function (PDF) of $\mathrm{v}_3 \theta$ is then computed for each of these frames. A time averaged PDF of local heat fluxes is then obtained using a minimum of 300 frames of computed data sets. Fig.~\ref{fig:Pr_4.0} shows  PDFs of the vertical local heat fluxes for $\mathrm{Ra} = 5.0 \times 10^5$ and $\mathrm{Pr} = 4.0$ for four different values of $\mathrm{Q}$ on the semi-log scale. Local heat fluxes are in the upward direction as well as in the downwards direction. All the PDFs are asymmetric about their maxima located at $\mathrm{v}_3 \theta = 0$ and are non-Gaussian. The asymmetry of the PDF shows that the excursion of upward heat flux is more than the excursion of downward heat flux. This signifies that a net heat flux is maintained in the vertically upward direction.  The data points in PDFs shown by blue (black) squares, pink (light gray) circles, magenta (gray) stars and green (gray) triangles are for $\mathrm{Q} = 70$, $50$, $300$ and $500$, respectively. 
The time averaged PDFs of local thermal fluxes in the vertical direction show a cusp at the maximum. This type of cusp was first observed in experiment on turbulent RBC~\cite{Shang_etal_prl_2003}. The PDF of instantaneous local fluxes in the vertical direction also showed the cusp at the maxima in simulations~\cite{Shishkina_Wagner_2007}. It may be due to non-Gaussian nature of the vertical velocity $\mathrm{v}_3$  and the convective temperature $\theta$.
The inset of Fig.~\ref{fig:Pr_4.0} shows an enlarged view of the PDFs near their maxima. The time averaged PDFs with wider spread have lower values of maxima. For $\mathrm{Ra} = 5.0 \times 10^5$ and $\mathrm{Pr} = 4.0$, the largest spread of a PDF is for $\mathrm{Q} = 70$. The histograms for these cases have exactly the similar shapes (not shown here) and they show the time averaged vertical local heat flux is maximum  for $\mathrm{Q} = 70$, which correspond to $\sqrt{\mathrm{Ra}/\mathrm{(Q Pr)}} = 42.25$ for $\mathrm{Pr} = 4.0$.  This is consistent with the largest global heat flux for $\mathrm{Q} = 70$ for the same set of all parameters (see Fig.~\ref{fig:Nu_Q}). The probability distribution functions of the local heat fluxes show exponential tails.  A large part of the distribution function for the upward local heat flux may be represented with two exponential functions, while the distribution function for the downward local heat flux can be represented well by a single exponential function.  It is interesting to note that local energy flux shows approximately exponential tails in wave turbulence~\cite{falcon_etal_prl_2008}. However, the anisotropy in a thermally stratified system makes the probability distributions of local heat fluxes in vertical direction asymmetric on two sides of the peak.

\section{\label{sec:conclusions}Conclusions}
                                                                                                                                                                                                                                                                                                                                                                                                                                                                                                 
A numerical study on global as well as local heat fluxes in Rayleigh-B\'{e}nard magnetoconvection in different fluids is presented. The global heat flux of unsteady magnetoconvection with a uniform vertical magnetic field shows a mild enhancement as the strength of the uniform magnetic field is raised for relatively lower values of $\mathrm{Q}$ and a range of thermal Prandtl number ($0.1 \le \mathrm{Pr} \le 4.0$). For relatively higher values of external magnetic field,  there is suppression of heat flux in nanofluids, liquid crystals as well in geophysical liquid metals. The time averaged global heat flux (Nusselt number) decreases logarithmically with Chandrasekhar's number for all fluids responsive to a vertical magnetic field, when Rayleigh number and Prandtl numbers are kept at fixed values.  For water based nanofluids is likely to show this behaviour, if the volume fraction of spherical copper nanoparticles  is around $8\%$. A similar behaviour is likely in Earth's outer liquid core ($\mathrm{Pr} \approx 0.1$) as well as some liquid crystals. The enhancement in heat flux makes the relative time averaged value of Nusselt number slightly more than unity for smaller values of the dimensionless parameter $\sqrt{\mathrm{Ra/(Q Pr)}}$.  The global thermal flux as well as the PDF of local heat flux confirm small enhancement of thermal flux. The fluctuating part of the Nusselt number shows nearly normal distribution with asymmetric tails. The power spectral density of the Nusselt number scales with frequency $f$ approximately as $f^{-2}$ for higher values of $f$. The thickness of thermal boundary layer scales with Rayleigh number as $\delta_{th} \sim \mathrm{Ra}^{-\gamma}$. For lower values of $\mathrm{Q}$,  $\gamma \approx 0.3$ and its value decreases as Chandrasekhar numbers is increased. The PDF of vertical local heat-fluxes is found to be non-Gaussian and asymmetric with  cusp at its maximum and it has exponential tails.\\

\noindent {\bf Acknowledgments:}\\
We thank both the anonymous Reviewers whose comments made us improve the manuscript significantly. Discussions with Dr. H.K. Pharasi was fruitful.
 

\end{document}